\documentclass[dvips]{acta}
\usepackage{supertabular,lscape,epsfig}
\usepackage{amssymb}
\usepackage{amsmath}
\usepackage{wasysym}
\DeclareSymbolFont{ppa}{OT1}{ppl}{m}{it}
\DeclareMathSymbol{\vv}{\mathalpha}{ppa}{'166}

\SetPages{1}{18}

\SetVol{59}{2009}

\begin{document}

\newcommand{\TabCapp}[2]{\begin{center}\parbox[t]{#1}{\centerline{
  \small {\spaceskip 2pt plus 1pt minus 1pt T a b l e}
  \refstepcounter{table}\thetable}
  \vskip2mm
  \centerline{\footnotesize #2}}
  \vskip3mm
\end{center}}

\newcommand{\TTabCap}[3]{\begin{center}\parbox[t]{#1}{\centerline{
  \small {\spaceskip 2pt plus 1pt minus 1pt T a b l e}
  \refstepcounter{table}\thetable}
  \vskip2mm
  \centerline{\footnotesize #2}
  \centerline{\footnotesize #3}}
  \vskip1mm
\end{center}}

\newcommand{\MakeTableSepp}[4]{\begin{table}[p]\TabCapp{#2}{#3}
  \begin{center} \TableFont \begin{tabular}{#1} #4
  \end{tabular}\end{center}\end{table}}

\newcommand{\MakeTableee}[4]{\begin{table}[htb]\TabCapp{#2}{#3}
  \begin{center} \TableFont \begin{tabular}{#1} #4
  \end{tabular}\end{center}\end{table}}

\newcommand{\MakeTablee}[5]{\begin{table}[htb]\TTabCap{#2}{#3}{#4}
  \begin{center} \TableFont \begin{tabular}{#1} #5
  \end{tabular}\end{center}\end{table}}

\newfont{\bb}{ptmbi8t at 12pt}
\newfont{\bbb}{cmbxti10}
\newfont{\bbbb}{cmbxti10 at 9pt}
\newcommand{\uprule}{\rule{0pt}{2.5ex}}
\newcommand{\douprule}{\rule[-2ex]{0pt}{4.5ex}}
\newcommand{\dorule}{\rule[-2ex]{0pt}{2ex}}
\def\thefootnote{\fnsymbol{footnote}}

\hyphenation{Ce-phe-ids eclip-ses chan-ges me-thod}

\begin{Titlepage}
\Title{The Optical Gravitational Lensing Experiment.\\
The OGLE-III Catalog of Variable Stars.\\
III. RR~Lyrae Stars in the Large Magellanic Cloud\footnote{Based on
observations obtained with the 1.3-m Warsaw telescope at the Las Campanas
Observatory of the Carnegie Institution of Washington.}}
\Author{I.~~S~o~s~z~y~\'n~s~k~i$^1$,~~
A.~~U~d~a~l~s~k~i$^1$,~~
M.\,K.~~S~z~y~m~a~\'n~s~k~i$^1$,\\
M.~~K~u~b~i~a~k$^1$,~~
G.~~P~i~e~t~r~z~y~\'n~s~k~i$^{1,2}$,~~
\L.~~W~y~r~z~y~k~o~w~s~k~i$^3$,\\
O.~~S~z~e~w~c~z~y~k$^2$,
~~K.~~U~l~a~c~z~y~k$^1$~~
and~~R.~~P~o~l~e~s~k~i$^1$}
{$^1$Warsaw University Observatory, Al.~Ujazdowskie~4, 00-478~Warszawa, Poland\\
e-mail:
(soszynsk,udalski,msz,mk,pietrzyn,kulaczyk,rpoleski)@astrouw.edu.pl\\
$^2$ Universidad de Concepci{\'o}n, Departamento de Fisica, Casilla 160--C,
Concepci{\'o}n, Chile\\
e-mail: szewczyk@astro-udec.cl\\
$^3$ Institute of Astronomy, University of
Cambridge, Madingley Road, Cambridge CB3 0HA, UK\\
e-mail: wyrzykow@ast.cam.ac.uk}
\Received{March 16, 2009}
\end{Titlepage}
\Abstract{The third part of the OGLE-III Catalog of Variable Stars
comprises 24\,906 RR~Lyr stars in the Large Magellanic Cloud (LMC). This
sample consists of 17\,693 fundamental-mode (RRab), 4958 first-overtone
(RRc), 986 double-mode (RRd) and 1269 suspected second-overtone (RRe)
pulsators. 66 objects are foreground Galactic RR~Lyr stars. The catalog
data include basic photometric and astrometric properties of these RR~Lyr
stars, multi-epoch {\it VI} photometry and finding charts.

We detected one new RR~Lyr star with additional eclipsing variations. The
spatial distribution of RR~Lyr stars in the LMC is distinctly non-spherical
and it is elongated in the same direction as the LMC bar. The basic
statistical features of RR~Lyr stars in the LMC are provided. The apparent
{\it V}-band magnitudes for RRab stars have the modal value at 19.36~mag,
and for overtone RR~Lyr stars it is about 19.32~mag. The mean periods for
RRab, RRc and RRe stars are 0.576, 0.337 and 0.270~days,
respectively.}{Stars: variables: RR~Lyrae -- Stars: oscillations -- Stars:
Population II -- Magellanic Clouds}

\Section{Introduction}
RR~Lyr stars are radially pulsating stars with periods in the range between
0.2 and 1.0~day. These are old, relatively low mass stars populating the
horizontal branch in the HR diagram. The role of RR~Lyr in modern
astrophysics cannot be overestimated. They are used for tracing the
chemical and dynamical properties of old stellar populations in our and
nearby galaxies. RR~Lyr variables are one of the cornerstones of the
astronomical distance scale. Finally, they are used as test objects for
evolutionary and pulsation models of low-mass stars.

Probably, the first known RR~Lyr star was a field variable U~Leporis
discovered by Kapteyn (1890, see the discussion in Smith 1995). However,
within the next years hundreds of variable stars of the same type were
detected in globular clusters (Pickering and Bailey 1895), therefore the
whole class was referred to as ``cluster-type variables'' at that time. The
name ``RR~Lyr stars'' (from the brightest member of the class) was
officially accepted by the International Astronomical Union in 1948.

The first RR~Lyr variables in the Large Magellanic Cloud (LMC) were
identified by Thackeray and Wesselink (1953). This discovery confirmed the
Baade's (1952) large revision of the astronomical distance scale and proved
the existence of the Population~II component in the LMC. Most of the
surveys for RR~Lyr variables in the LMC undertaken during the subsequent
decades covered old clusters and their vicinities (\eg Alexander 1960,
Wesselink 1971, Graham and Ruiz 1974, Graham 1985, Nemec \etal 1985, Hazen
and Nemec 1992, Walker 1992).

The number of known RR~Lyr stars in the LMC was multiplied thanks to the
photometric databases collected by the large microlensing surveys: MACHO
and OGLE. Alcock \etal (1996) announced the discovery of about 7900 RR~Lyr
stars in the LMC on the basis of the MACHO observations. The OGLE-II
project has cataloged 7612 RR~Lyr stars (Soszyñski \etal 2003) detected in
the 4.5 square degrees of the central regions of the LMC. Deep photometric
and spectroscopic surveys for RR~Lyr stars, but covering limited regions in
the LMC, were conducted by Clementini \etal (2003), Borissova \etal (2004,
2006) and Di Fabrizio \etal (2005).

In this work we describe the catalog of nearly 25\,000 RR~Lyr stars in the
LMC. Almost 1000 of them are double-mode pulsators. This is the third part
of the OGLE-III Catalog of Variable Stars (OIII-CVS) -- the catalog which
is planned to comprise practically all variable sources in the OGLE-III
fields in the Magellanic Clouds and the Galactic bulge. In the previous
papers of this series we presented the catalogs of 3361 classical Cepheids
(Soszyñski \etal 2008a, hereafter Paper~I), 197 type II Cepheids and 83
anomalous Cepheids in the LMC (Soszyñski \etal 2008b, hereafter Paper~II).

This paper is organized as follows. In Section~2 we describe the reductions
and calibrations of the data. In Section~3 details on the RR~Lyr stars
identification and classification are provided. In Section~4 we describe
the catalog itself and compare it with other catalogs of RR~Lyr stars in
the LMC. In Section~5 we discuss some aspects concerning statistical
features of RR~Lyr stars in the LMC. In Section~6 we draw our conclusions.

\Section{Observations and Data Reduction}
Photometric observations presented in this catalog were carried out
with the 1.3-m Warsaw telescope located at Las Campanas Observatory,
Chile. The observatory is operated by the Carnegie Institution of
Washington. The ``second generation'' camera uses eight SITe
$2048\times4096$ CCD detectors with 15~$\mu$m pixels resulting in
0.26~arcsec/pixel scale and $35\arcm\times35\zdot\arcm5$ field of view. For
the details of the instrumentation setup we refer the reader to Udalski
(2003).

OGLE-III fields in the LMC cover nearly 40 square degrees and about
32~million stars. Approximately 400 photometric points per star were
accumulated over seven seasons, between July 2001 and March 2008. Most of
the observations were done through the Cousin's {\it I}-band filter with
exposure time 180~s. A~few dozen observations per star have been obtained
with the Johnson's {\it V}-band and integration time 225~s. The photometry
of stars in the central regions of the LMC is supplemented by the OGLE-II
data collected between 1997 and 2000 using the same Warsaw telescope. For
each individual star their mean magnitudes were derived independently for
the OGLE-II and OGLE-III datasets, and the OGLE-II photometry was shifted
to agree with the OGLE-III data.

Data reduction pipeline was based on the Difference Image Analysis (DIA --
Alard and Lupton 1998, Alard 2000, Wo¼niak 2000). Photometric errors
produced by the DIA package were corrected using a program developed by
J.~Skowron (the technique is described in detail in Wyrzykowski \etal
2009). Full description of the reduction techniques, photometric
calibration and astrometric transformations can be found in Udalski \etal
(2008).

\vskip7pt
\Section{Identification and Classification of RR~Lyr Stars}
\vskip3pt
\subsection{Single-Mode RR~Lyr Stars}
\vskip3pt
All light curves in the LMC collected during the OGLE-III project have
passed through a period search algorithm using supercomputers at the
Interdisciplinary Centre for Mathematical and Computational Modelling
(ICM). We used program {\sc Fnpeaks} (Ko³aczkowski, private communication)
which is based on the Fourier analysis. The frequency space ranged from 0
to 24 cycles per day, with a search interval of 0.0001 cycles per day. For
each star the primary period was derived, then the light curve was
pre-whitened with this period and the procedure of period search was
repeated on the residual data.

From the sample of 32 million stars in the LMC we filtered out objects with
signal-to-noise ratio of the dominant frequency smaller than 5. Then, we
subjected for visual inspection all stars with main periods between 0.2 and
1.0 day and brighter than $I=20$~mag (during further analysis we also
examined fainter objects but using higher threshold for the signal-to-noise
parameter).

As a result of the visual inspection we removed artefacts, obvious
eclipsing binaries and other non-pulsating stars. In case of doubts about
the proper classification we used colors, amplitudes and Fourier parameters
of the light curve decomposition to compare a given star with the whole
sample of RR~Lyr variables. However, in a number of cases we were not able
to judge the types of variable stars, especially among overtone RR~Lyr
stars which typically exhibited symmetric light curves, similar to close
eclipsing binaries. In the catalog we flagged these objects as uncertain.

During the classification process we detected 11 new classical Cepheids,
which were overlooked in Paper~I. We also noticed three extremely
short-period 1O/2O double-mode Cepheids (or $\delta$~Sct stars, depending
on the definition) with the first-overtone periods as short as
0.22~days. We included all these objects in the OGLE-III catalog of
classical Cepheids in the LMC (Paper~I) increasing the whole number of
objects in that catalog to 3375. In the short-period domain several dozens
of stars were categorized as high amplitude $\delta$~Sct (HADS) variables,
as they were located on the extension of the Cepheid period--luminosity
relation. Some of these objects exhibited secondary periods and followed
the sequence of F/1O double-mode HADS in the Petersen diagram. These
objects will be presented in one of the next parts of the OIII-CVS.

\begin{figure}[htb]
\centerline{\includegraphics[width=11.3cm, bb=10 400 585 745]{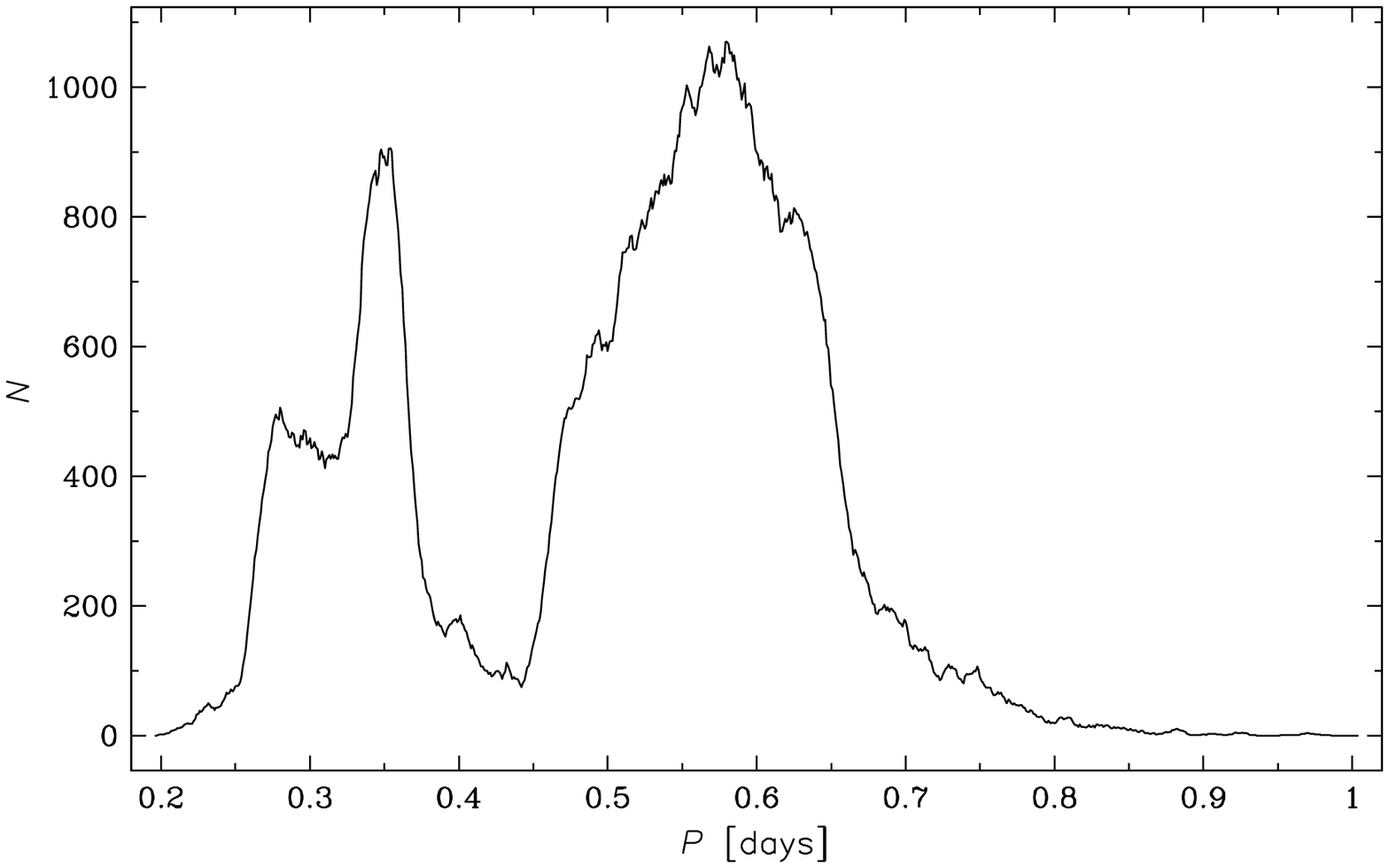}}
\FigCap{Period distribution of RR~Lyr stars in the LMC. The function is
composed of ten histograms (with the bin width equal of 0.01~days) shifted
by 1/10 of the bin width with respect to each other.}
\end{figure}
Our final sample of RR~Lyr stars consists of 24\,906 objects. This is the
largest set of RR~Lyr stars detected so far in any environment. The period
distribution for the entire sample is shown in Fig.~1. This plot is
composed of ten histograms (with the bin width of 0.01~days) shifted by
1/10 of the bin width with respect to each other. In such a way we could
visualize more subtle effects of the period distribution.

\begin{figure}[p]
\centerline{\includegraphics[width=13.5cm]{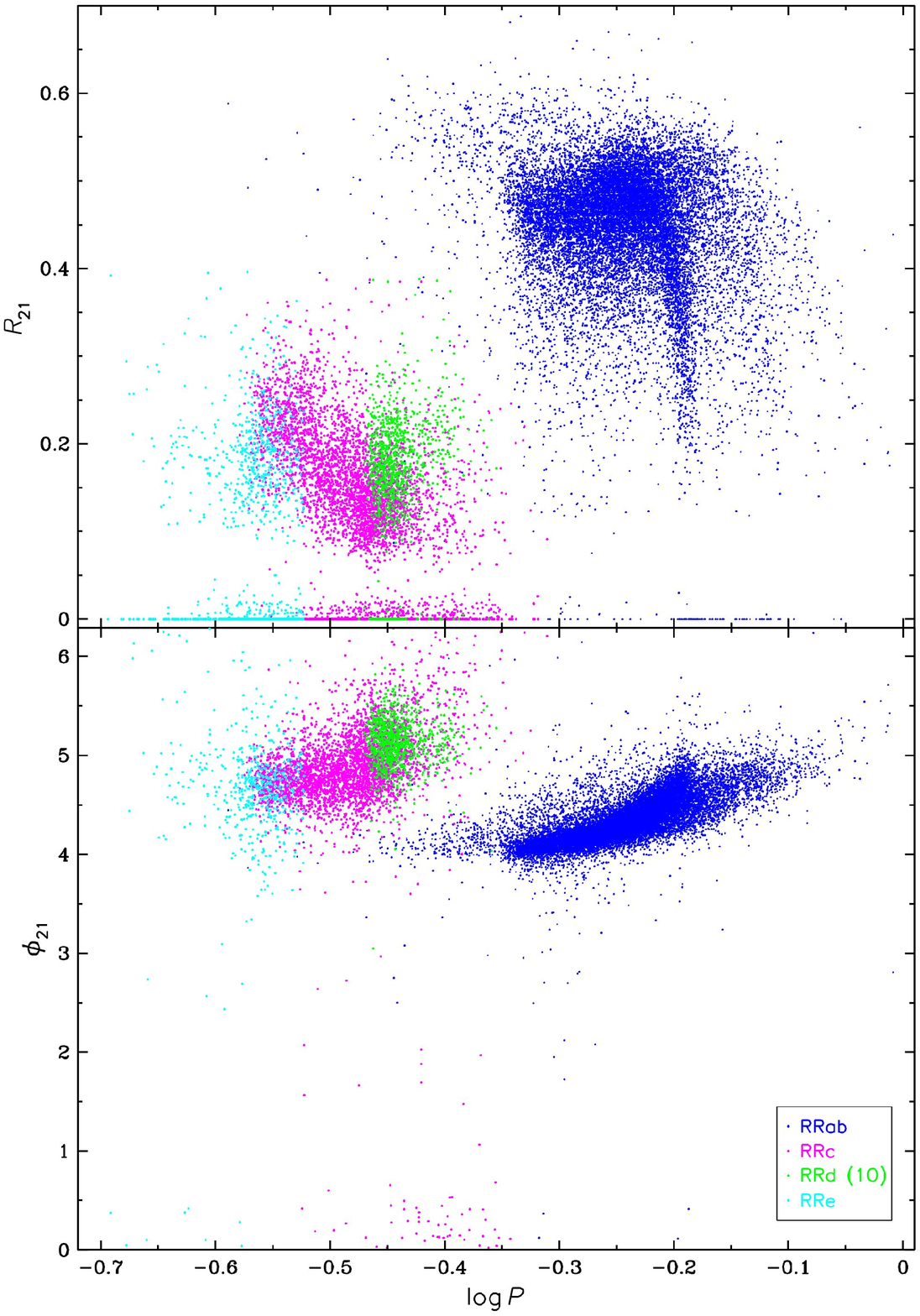}}
\FigCap{Fourier parameters $R_{21}$ and $\phi_{21}$ \vs $\log{P}$ for
RR~Lyr in the LMC. Blue, magenta, green and cyan points show RRab, RRc, RRd
(first overtone) and RRe stars, respectively.}
\end{figure}
In Fig.~1 we identify three well-known maxima at periods of about 0.58,
0.35 and 0.28~days. The first two peaks correspond to fundamental-mode
(RRab or RR0) and first-overtone (RRc or RR1) RR~Lyr stars. We
distinguished between both classes using shapes of their light curves that
can be quantitatively described with the parameters of the Fourier
decomposition: amplitude ratios $R_{k1}=A_k/A_1$ and phase differences
$\phi_{k1}=\phi_k-k\phi_1$ (Simon and Lee 1981). Fig.~2 presents Fourier
parameters $R_{21}$, $\phi_{21}$ plotted against $\log{P}$. The number of
harmonics of the Fourier decomposition in each star was adjusted to
minimize the $\chi^2$ per degree of freedom. In some stars, mainly
first-overtone pulsators, such solution gave only a pure sinusoid fit, and
obviously for these objects $R_{21}$ is equal to zero, while $\phi_{21}$ is
not defined.

As can be noticed in Fig.~2, RRab and RRc stars are well separated in the
diagrams showing the Fourier coefficients \vs periods. Thus, we used these
planes to divide RR~Lyr stars into fundamental-mode and overtone
pulsators. A number of objects lying close to the boundary between both
classes were visually examined and, in some cases, the automated
classification was changed.

The origin of the shortest-period maximum in the period distribution is
ambiguous. Alcock \etal (1996) suggested that such an additional peak may
be a signature of the second overtone oscillations (RRe or RR2 stars). The
other possibility is that these short period variables belong to a more
metal-rich population of the first overtone RR~Lyr stars (Bono \etal
1997). Regardless of the cause of this excess in the short-period domain,
we designate these objects as RRe stars.

\begin{figure}[b]
\vspace{-1mm}
\hglue-4mm{\includegraphics[width=13.45cm]{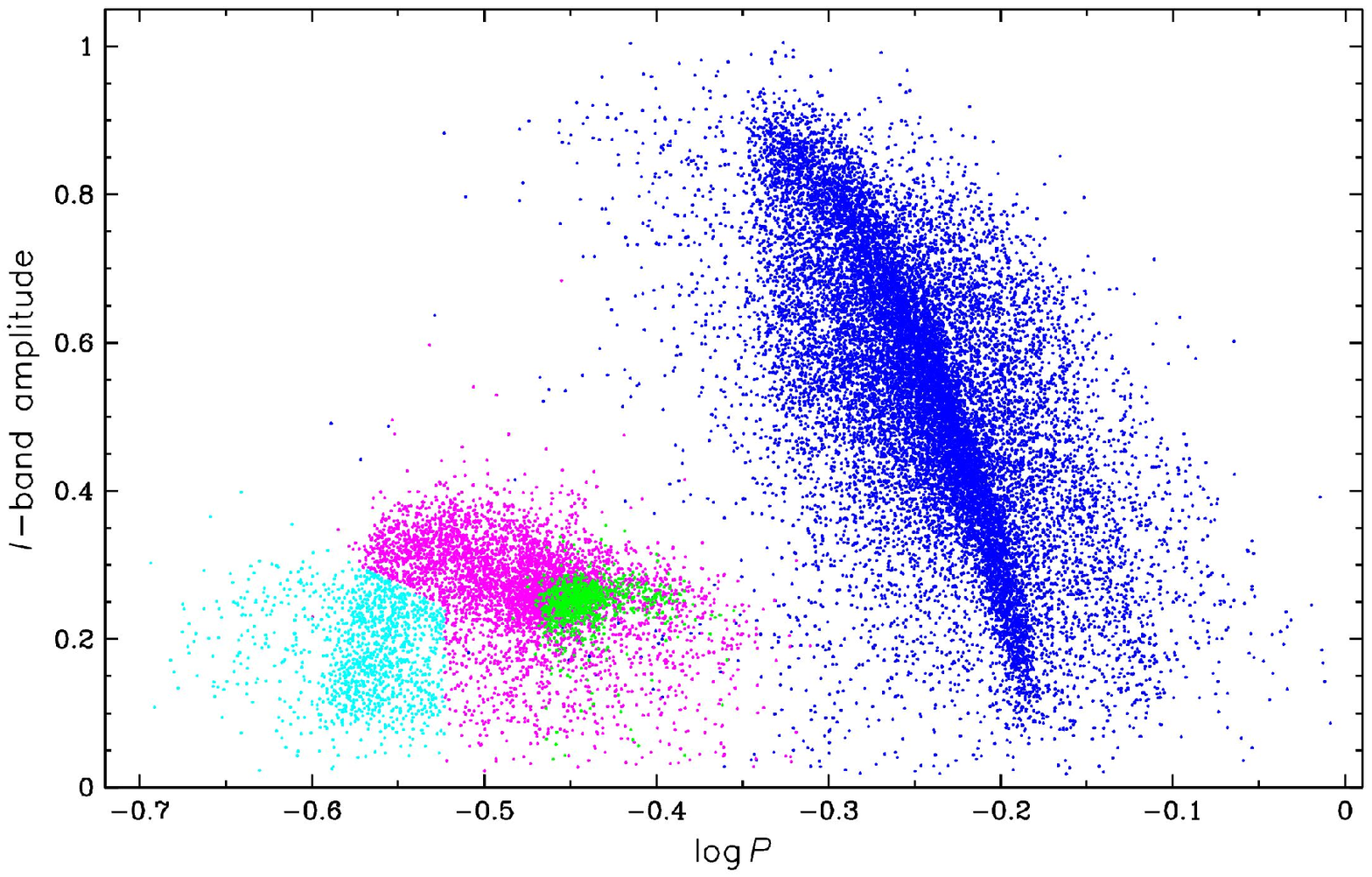}}
\FigCap{Period--amplitude diagram for RR~Lyr stars in the LMC. The color
symbols are the same as in Fig.~2.}
\end{figure}
We note that RRc and RRe variables overlap in all diagrams that can be
plotted using the OGLE-III data, so distinguishing between both groups is
correct only in a statistical sense. In individual cases it can be
wrong. To separate RRc and RRe stars we used period--amplitude diagram
(sometimes called the Bailey diagram) presented in Fig.~3. RR~Lyr stars
with periods shorter than 0.3~days were divided into two groups according
to their {\it I}-band amplitudes. Objects with lower amplitudes were
classified as RRe stars.

\subsection{Multi-Periodic RR~Lyr Stars}
Double-mode RR~Lyr stars, known as RRd or RR01 stars, exhibit simultaneous
oscillations in the fundamental and first overtone radial modes. Alcock
\etal (1997a, 2000) reported the discovery of 181 RRd stars in the LMC. In
the OGLE-II catalog of RR~Lyr stars in the LMC (Soszyñski
\etal 2003) 230 objects were classified as double-mode pulsators.

The search for multi-periodic RR~Lyr stars was carried out in two
ways. First, we used the database of periods derived for all stars in the
LMC. We selected the light curves that had statistically significant
primary and secondary periods and their position in the Petersen diagram
(\ie the plot of the period ratio \vs the longer period) was in agreement
with the region occupied by RRd stars, \ie the longer period ranged between
0.44 and 0.60~days and the period ratio was between 0.74 and 0.75. These
light curves were visually inspected and the initial list of double-mode
RR~Lyr stars was prepared.

The second part of the double-mode search was performed for all RR~Lyr
stars selected before. Each light curve was fitted with the Fourier series
and the residuals of the fit were searched for secondary frequencies.
Again, we visually examined the light curves with periods and period ratios
similar to these of RRd stars.

\begin{figure}[htb]
\centerline{\includegraphics[width=16cm, bb=20 160 575 745]{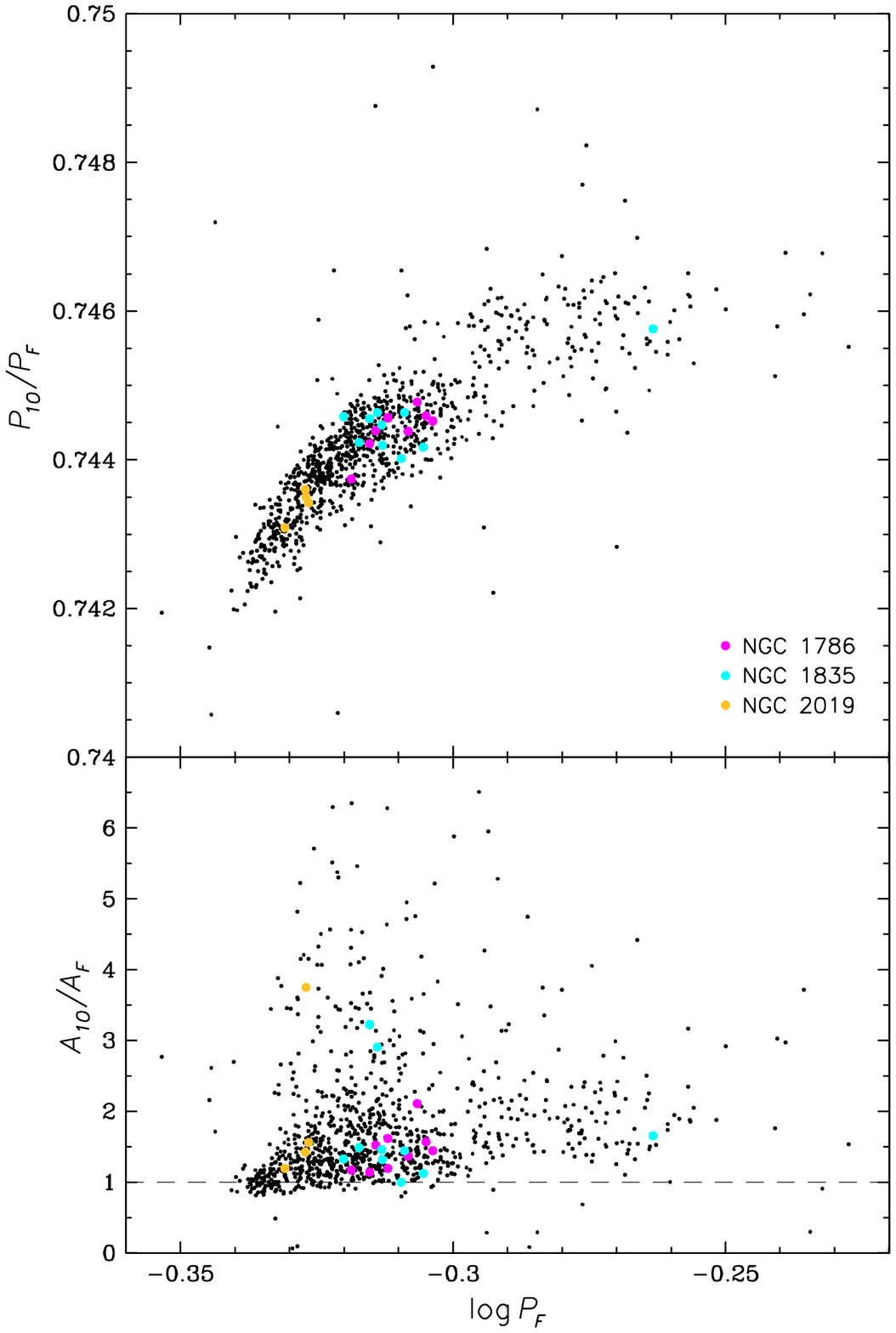}}
\FigCap{Petersen diagram for RRd stars in the LMC ({\it upper panel}) and
the amplitude ratio between first overtone and fundamental mode \vs
$\log{P}$ ({\it lower panel}). Color symbols mark RRd stars in the three
LMC globular clusters: NGC~1786 (magenta), NGC~1835 (cyan) and NGC~2019
(yellow).}
\end{figure}
Our final list of RRd variables contains 986 objects. Fig.~4 shows the
Petersen diagram for these stars. The well known curved sequence in this
diagram is clearly visible, although several outlying points also
exist. These stars may be RRd variables with unusual physical properties
(see Popielski \etal 2000), but it is also possible that the secondary
periods in these objects are caused by something else than the radial
pulsations (\eg nonradial oscillations, blending with another periodic
variable, etc.). Lower panel of Fig.~4 shows ratio of {\it I}-band
amplitudes in both modes (first overtone to fundamental mode) against
$\log{P}$. The first overtone dominates in most of the RRd stars, but in
some cases the fundamental mode variations have larger amplitudes, in
particular for stars with the shortest periods. It is evident that the
typical minimum and maximum values of the $A_{1O}/A_F$ amplitude ratios are
correlated with periods.

Apart from RRd stars, we found a number of double-periodic RR~Lyr variables
with ratios of periods inconsistent with simultaneous oscillations in the
fundamental mode and the first overtone. For about 20\% of RRab stars, 19\%
of RRc and 25\% of RRe stars we detected secondary frequencies very close
to the primary ones, with period ratios larger than 0.9. Such stars,
commonly referred to as Blazhko RR~Lyr stars, are suspected to exhibit
nonradial modes of pulsation (Olech \etal 1999, Dziembowski and Mizerski
2004). Alcock \etal (2000) distinguished several variants of this behavior,
with two, three or more close frequency components in the power
spectra. Our simple frequency analysis does not settle the question of
which variant is exhibited in a given object. Moreover, the secondary
periods in the residual data may also be produced by RR~Lyr stars changing
their periods. Temporal coverage of the OGLE photometric database is long
enough to study the period changes of RR~Lyr stars. For many objects,
mainly RRc stars, we noticed that their light curves cannot be folded with
the constant periods, because the rates of the period changes are so
large. Similar considerable period changes that cannot be explained by the
evolutionary effects were recently found in the LMC classical Cepheids by
Poleski (2008). The relatively large (compared to other investigations)
incident rate of RRc stars with close frequencies is likely caused by these
period-changing pulsators.

A number of RR~Lyr stars show secondary periodicities which are distinctly
different than primary ones, but these objects cannot be unambiguously
categorized to any group. For example, in two stars (OGLE-LMC-RRLYR-11983,
OGLE-LMC-RRLYR-14178) we detected secondary periods that give period ratios
of about 0.60--0.61. Similar objects were recently discovered by Olech and
Moskalik (2009) in $\omega$~Cen. Information about the secondary
periodicities for individual stars is provided in the remarks in the
catalog.

Olech and Moskalik (2009) also announced the discovery of a candidate
double-mode RR~Lyr star with the first and the second overtones excited. In
our sample we did not find any reliable candidate for such an
object. Admittedly, one of the RRd stars (OGLE-LMC-RRLYR-02746) shows an
additional period of 0.291121~days giving the period ratio of 0.806 with
the first-overtone mode, but we detected another RR~Lyr variable
(OGLE-LMC-RRLYR-02743) with the same period \linebreak(0.291121~days) and
located at a distance of only 0\zdot\arcs7 from the former star. Since this
distance is smaller than the typical size of the seeing disk in the
OGLE-III frames (1\zdot\arcs2), we suspect the tertiary period in
OGLE-LMC-RRLYR-02746 is an artefact produced by blending with the other
RR~Lyr star. The image taken by the Hubble Space Telescope and retrieved
from the Hubble archive indeed confirms that two close stars are present in
this location.

\begin{figure}[htb]
\centerline{\includegraphics[width=13.5cm, bb=70 590 545 755]{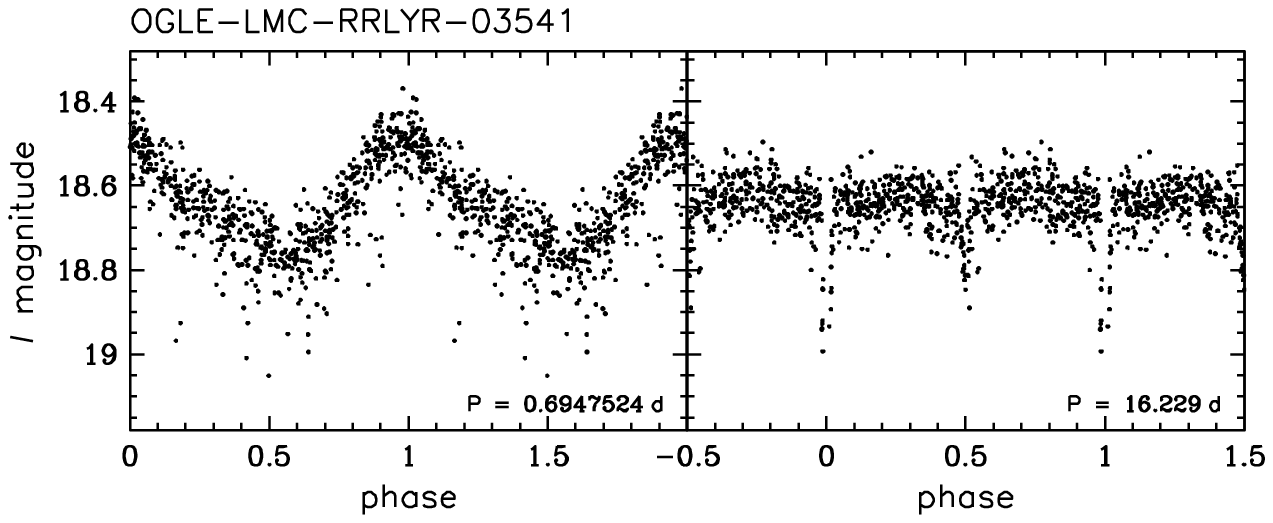}}
\FigCap{Light curve of the RR~Lyr star with additional eclipsing
variability. {\it Left panel}: the original photometric data folded with
pulsation period. {\it Right panel}: eclipsing light curve after
subtracting the RR~Lyr component.}
\end{figure}
A special attention was paid to search for RR~Lyr stars with additional,
eclipsing variations overimposed on the pulsation light curves. At present,
no RR~Lyr star being a member of an eclipsing binary system is known. The
OGLE-II project yielded three RR~Lyr stars with simultaneous eclipsing
modulation (Soszyñski \etal 2003), but all of these objects turned out to
be optical blends (Pr{\v s}a \etal 2008). In the present sample of RR~Lyr
stars we re-detected these three cases and found one additional RR~Lyr
variable with eclipses. OGLE-LMC-RRLYR-03541 has orbital period equal to
16.229~days and it is a good candidate for RR~Lyr star in an eclipsing
binary system. The light curve of this object is plotted in Fig.~5. The
original {\it I}-band photometry folded with the pulsation period is shown
in the left panel, while the right panel shows the eclipsing light curve
after subtracting the RR~Lyr component.

It is interesting that among 24\,906 RR~Lyr stars we found only one
candidate for eclipsing binary system with a pulsating star as one of the
components, while the catalog of 3361 classical Cepheids in the LMC
(Paper~I) contains four such objects (including an eclipsing system of two
Cepheids) and the catalog of 197 type~II Cepheids in the LMC (Paper~II)
lists as many as seven pulsating stars with eclipses. These extremely
different incident rates of binary systems among pulsating stars can be
explained by different evolution of these stars in the past. The radius of
a star at the tip of the red giant branch is by a factor of 15 larger than
the radius of a star located on the horizontal branch.  Thus, only
long-period binary systems (with periods longer than several hundred days;
Dziembowski, private communication) can avoid mass exchange in the phase of
the first ascent red giant branch. Now, in the RR~Lyr phase, the stars have
much smaller sizes and the probability of producing eclipses in such
long-period binary systems is very low.

Assuming that OGLE-LMC-RRLYR-03541 is a member of the binary system, its
orbital period is too short for the system to remain detached during the
previous phases of the stellar evolution. Explaining the evolution of such
a system could be a real challenge for the theory.

\Section{The Catalog}
The OGLE-III catalog of RR~Lyr stars in the LMC contains 24\,906 objects,
of which 17\,693 are RRab, 4958 -- RRc, 986 -- RRd, and 1269 -- RRe
stars. The list of objects with their multi-epoch {\it VI} photometry and
finding charts is available on-line through the WWW interface or {\it via}
anonymous FTP site:
\begin{center}
{\it http://ogle.astrouw.edu.pl/} \\
{\it ftp://ftp.astrouw.edu.pl/ogle/ogle3/OIII-CVS/lmc/rrlyr/}\\
\end{center}

In the FTP site the full list of RR~Lyr stars is given in the file {\sf
ident.dat}. The stars are listed in order of increasing right ascension and
designated with symbols OGLE-LMC-RRLYR-NNNNN, where NNNNN is a five digit
consecutive number. The file {\sf ident.dat} contains the following
information about each RR~Lyr star: the object designation, OGLE-III field
and internal database number of a star, mode of pulsation (RRab, RRc, RRd
or RRe), equinox J2000.0 right ascension and declination,
cross-identifications with the OGLE-II catalog of RR~Lyr stars in the LMC
(Soszyñski \etal 2003), with the MACHO catalog (Alcock \etal 1996, 1997a,
2000) and with the extragalactic part of the General Catalogue of Variable
Stars (GCVS -- Artyukhina \etal 1995). In the last column there are other
designations taken from the GCVS.

Basic parameters of the RR~Lyr stars are provided in the files {\sf
RRab.dat}, {\sf RRc.dat}, {\sf RRd.dat} and {\sf RRe.dat}. For single-mode
objects the consecutive columns contain: object designation, intensity mean
magnitudes in the $I$ and $V$ bands, periods in days and their
uncertainties, epochs of maximum light, peak-to-peak {\it I}-band
amplitudes, and Fourier parameters $R_{21}$, $\phi_{21}$, $R_{31}$,
$\phi_{31}$ derived for the {\it I}-band light curves. For RRd stars the
format of the table is longer including secondary periodicities. Periods
and their uncertainties were derived using program {\sc Tatry} by
Schwarzenberg-Czerny (1996).

The file {\sf remarks.txt} contains additional information on some RR~Lyr
stars. The subdirectory {\sf phot/} contains multi-epoch {\it I}- and {\it
V}-band OGLE photometry of the stars. If available, OGLE-II data are merged
with the OGLE-III photometry. The subdirectory {\sf fcharts/} contains
finding charts of all objects. These are the $60\arcs\times60\arcs$
subframes of the {\it I}-band DIA reference images, oriented with N up, and
E to the left.

To test the completeness of our catalog we matched our initially selected
sample with two large selections of RR~Lyr stars in the LMC -- the OGLE-II
and the MACHO catalogs. The OGLE-II project released the catalog of 7612
RR~Lyr variables (Soszyñski \etal 2003). In the present sample, we found no
counterparts for 205 of them. From that number 50 stars were reclassified
in the present investigation as classical or anomalous Cepheids, HADS or
eclipsing binary systems. Most of the remaining 155 stars were located
close to the edges of the OGLE-III fields and were affected by a small
number of observations. We included these missing objects in the present
catalog.

The list of RR~Lyr stars in the LMC detected by the MACHO project was
retrieved from the web-page of the
survey\footnote{http://wwwmacho.mcmaster.ca/}. After removing double
detections and stars lying outside the OGLE-III fields we found 8745 stars
in the list. In the preliminary version of our catalog we missed 151 of
them. However, only 39 of these missing stars seem to be RR~Lyr
variables. The remaining objects usually belong to different types of
variable stars, most of them being eclipsing binaries. It is worth noting
that for about 220 positively cross-identified RR~Lyr stars the periods
provided in the MACHO list seemed to be aliases of the periods derived
here.

\begin{figure}[p]
\centerline{\includegraphics[width=12.15cm]{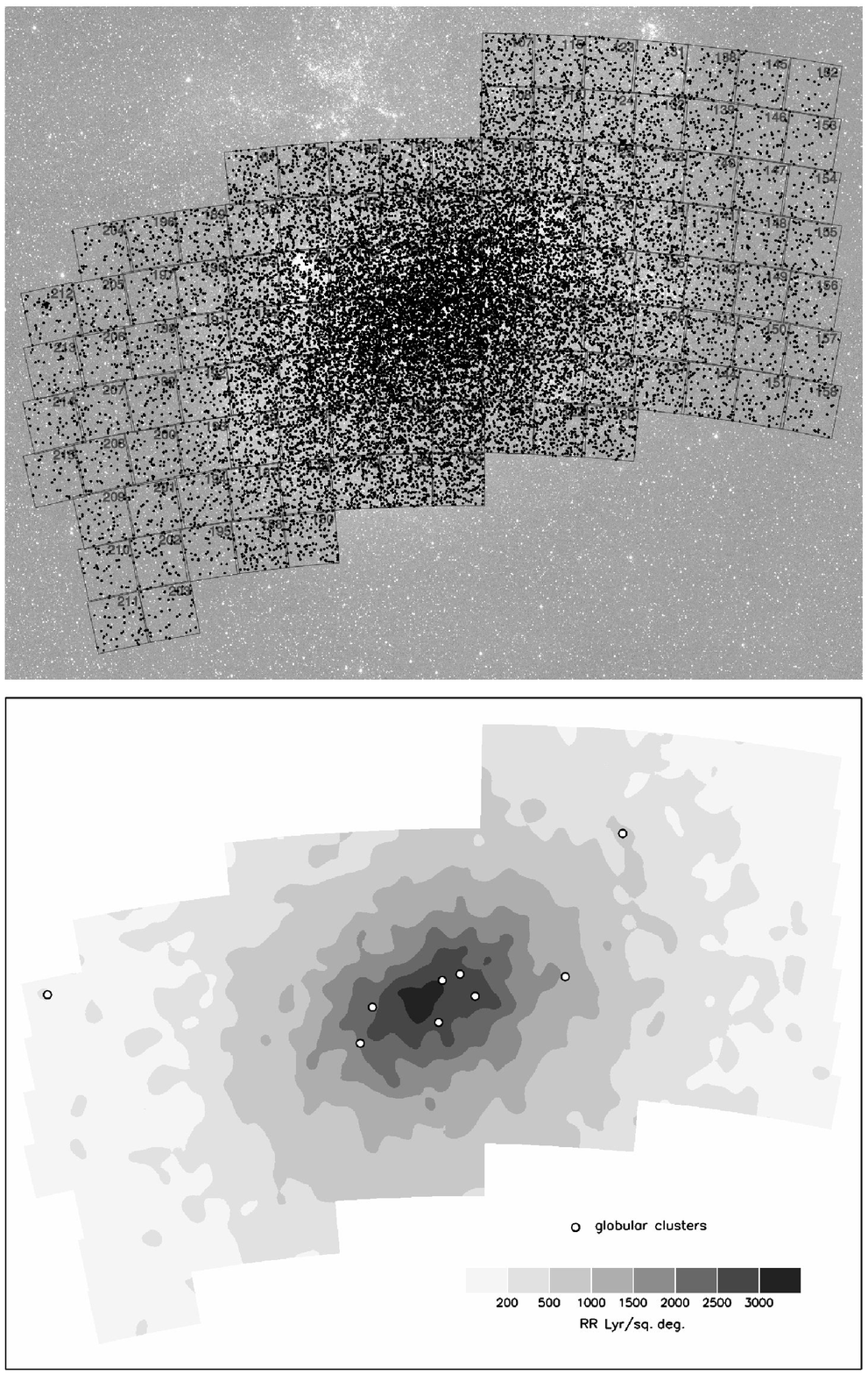}}
\vspace{1mm}
\FigCap{{\it Upper panel}: spatial distribution of RR~Lyr stars in the
LMC. The background image of the LMC is originated from the ASAS sky
survey. {\it Lower panel}: surface density map of RR~Lyr stars in the
LMC. White circles show positions of globular clusters.}
\end{figure}
Our sample was also compared with the LMC RR~Lyr stars listed in the GCVS
(Artyukhina \etal 1995). Unfortunately, 151 of the 197 LMC stars classified
as RR~Lyr stars in the GCVS were located outside the OGLE-III fields. From
the remaining variables our catalog includes 42 objects. Two stars
(LMC~V0588, LMC~V4498) turned out to be classical Cepheids (Paper~I), one
(LMC~V0464) was classified as an anomalous Cepheid (Paper~II). We found no
counterpart for one star (LMC~V0546) classified in the GCVS as "RRAB:".

\Section{Discussion}
The upper panel of Fig.~6 displays the position of OGLE-III RR~Lyr stars
overplotted on the image originated in the ASAS-3 survey (Pojmañski
1997). After smoothing this distribution with the Gaussian filter
we obtained a surface density map visualized in the lower panel of
Fig.~6. It is evident that the distribution of field RR~Lyr stars in the
LMC is elongated along the LMC bar. This result confirms the conclusion of
Subramaniam (2006) drawn on a basis of the OGLE-II catalog of RR~Lyr
stars.

In the lower panel of Fig.~6 we also marked globular clusters in which
RR~Lyr stars were found. The catalog of extended objects in the Magellanic
System recently published by Bica \etal (2008) lists 16 genuine globular
clusters in the LMC. Five of these objects are out of the OGLE-III
fields. We used the coordinates and angular sizes for the remaining 11
clusters provided by Bica \etal (2008) to select RR~Lyr stars located
inside the area outlined by the cluster radii.

\MakeTableee{
c@{\hspace{8pt}} c@{\hspace{6pt}} c@{\hspace{6pt}} c@{\hspace{8pt}}
c@{\hspace{6pt}} c@{\hspace{6pt}} c@{\hspace{6pt}} c@{\hspace{6pt}}
c@{\hspace{2pt}}}{12.5cm}{Globular clusters containing RR~Lyr stars}{\hline
\noalign{\vskip3pt}
Cluster & RA & Dec & Cluster & $N_{\rm RR}$ & $N_{\rm ab}$ & $N_{\rm c}$ & $N_{\rm d}$ & $N_{\rm e}$ \\
 name & (J2000) & (J2000) & radius [\arcm] & & & & & \\
\noalign{\vskip3pt}
\hline
\noalign{\vskip3pt}
 NGC~1754 & 4\uph54\upm17\ups & $-70\arcd26\arcm29\arcs$ & 1.6 & 36 & 20 & 15 & 0 & 1 \\
 NGC~1786 & 4\uph59\upm06\ups & $-67\arcd44\arcm42\arcs$ & 2.0 & 55 & 28 & 18 & 9 & 0 \\
 NGC~1835 & 5\uph05\upm06\ups & $-69\arcd24\arcm14\arcs$ & 2.3 & 109 & 63 & 30 & 10 & 6 \\
 NGC~1898 & 5\uph16\upm41\ups & $-69\arcd39\arcm23\arcs$ & 1.6 & 49 & 31 & 16 & 0 & 2 \\
 NGC~1916 & 5\uph18\upm38\ups & $-69\arcd24\arcm23\arcs$ & 2.1 & 25 & 15 & 10 & 0 & 0 \\
 NGC~1928 & 5\uph20\upm57\ups & $-69\arcd28\arcm40\arcs$ & 1.3 &  8 & 7 & 1 & 0 & 0 \\
 NGC~1939 & 5\uph21\upm26\ups & $-69\arcd56\arcm59\arcs$ & 1.4 &  7 & 3 & 4 & 0 & 0 \\
 NGC~2005 & 5\uph30\upm10\ups & $-69\arcd45\arcm10\arcs$ & 1.6 & 18 & 9 & 9 & 0 & 0 \\
 NGC~2019 & 5\uph31\upm56\ups & $-70\arcd09\arcm33\arcs$ & 1.5 & 61 & 36 & 15 & 4 & 6 \\
 NGC~2210 & 6\uph11\upm31\ups & $-69\arcd07\arcm18\arcs$ & 3.3 & 58 & 34 & 21 & 0 & 3 \\
\noalign{\vskip3pt}
\hline}
In the cluster Hodge~11 we found only one RR~Lyr variable and it is
probably a field star. Toward the clusters NGC~1939 and NGC~1928 we
detected seven and eight RR~Lyr stars, respectively. These numbers are
somewhat larger than the expected numbers of field RR~Lyr stars in the
regions occupied by clusters, but the possibility that all these variables
belong to a field cannot be ruled out. For the remaining eight globular
clusters there are no doubts that the bulk of RR~Lyr stars identified
around them belong to these clusters. Table~1 summarizes the information
about RR~Lyr stars in the star clusters in the LMC. This table lists, from
the left to right, designation of the cluster, coordinates of the cluster
center and cluster radius (from Bica \etal 2008), the total number of
RR~Lyr stars detected within the cluster radius and the number of RRab,
RRc, RRd and RRe stars in each cluster.

Three of the clusters presented in Table~1 host RRd stars. In Fig.~4 we
highlighted these objects in colors to show that the cluster RRd stars
occupy relatively limited area in the Petersen diagram compared to the
field variables. This is related to a much smaller range of metal
abundances in the LMC clusters than in the field.

The period--luminosity (PL) diagrams for RR~Lyr stars in the LMC in $V$,
$I$ and extinction insensitive Wesenheit index $W_I=I-1.55(V-I)$ are
plotted in Fig.~7. The magnitudes shown in the two upper panels are not
compensated for interstellar extinction. The PL relations are clearly
visible with the slopes which depend on the waveband. One can notice that
many objects in our catalog are significantly brighter or fainter than
typical RR~Lyr stars with a given period. Most of the RR~Lyr stars above
the PL relations (in case of the $\log{P}$--$W_I$ diagram also below the
relation) are blended objects. Their amplitudes are usually reduced
compared to unblended RR~Lyr variables with the same periods.
\begin{figure}[p]
\centerline{\includegraphics[width=13.41cm]{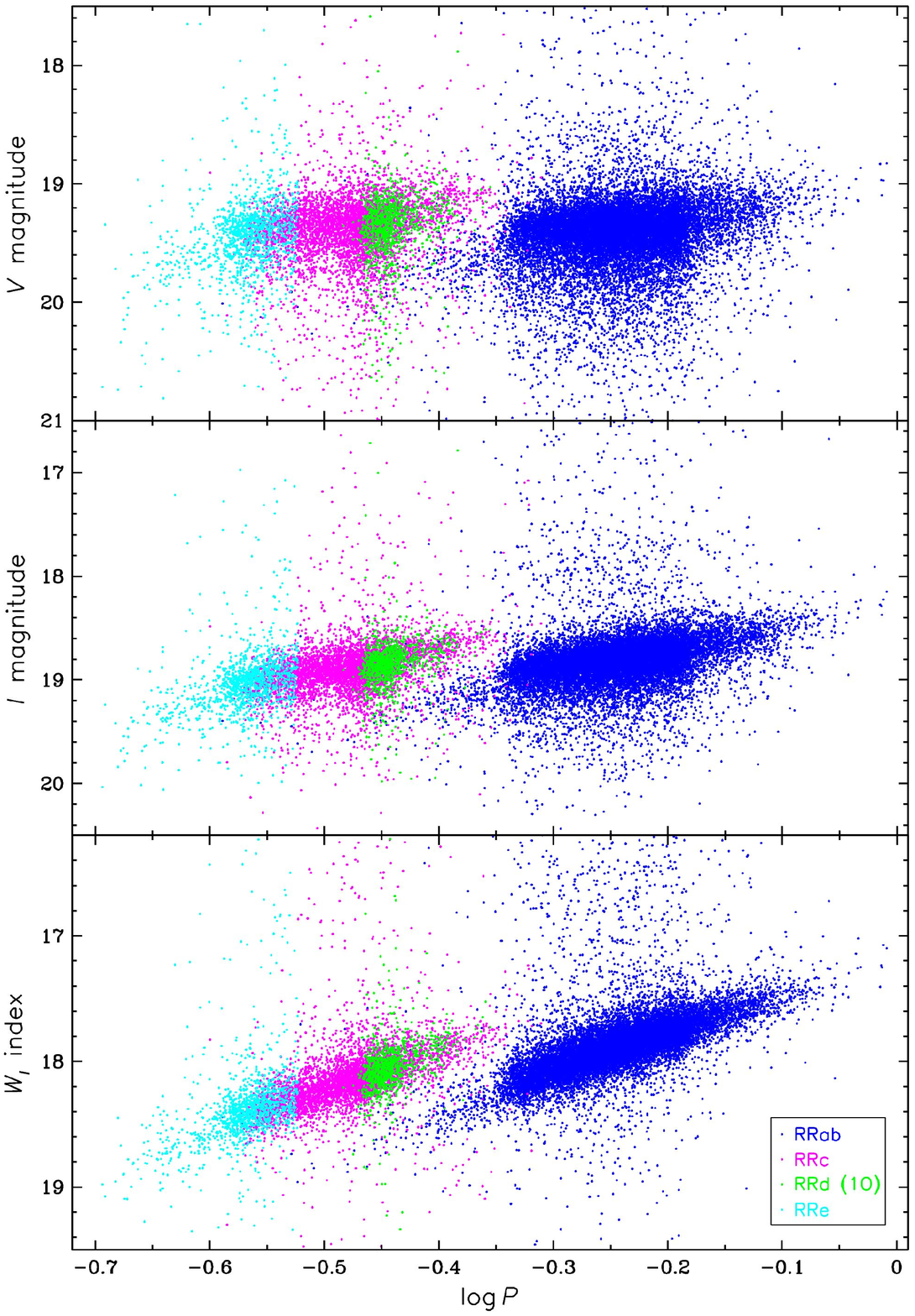}}
\FigCap{Period--luminosity diagrams for RR~Lyr stars in the LMC. The color
symbols are the same as in Fig.~2.}
\end{figure}

However, among these bright stars we detected 66 Galactic RR~Lyr stars
lying in the front of the LMC. Their light curves, amplitudes and $(V-I)$
colors are similar to the LMC RR~Lyr stars. 55 of these foreground RR~Lyr
stars are RRab variables, 8 were classified as RRc stars, 1 as RRe, and 2
of them are double mode RRd stars. The distances to the Galactic RR~Lyr
stars derived on the basis of their magnitudes seem to be almost uniformly
distributed, with no concentration toward any particular distance between
us and the LMC. It is in agreement with the result presented by Alcock
\etal (1997b) who used the MACHO database to select 20 RR~Lyr stars lying
in the front of the LMC.

The bulk of RR~Lyr stars significantly fainter than the PL relations are
highly reddened objects. They are well visible in the color--magnitude
diagram plotted in Fig.~8. The tail of points in the lower right part of the
diagram outlines the reddening vector.
\begin{figure}[htb]
\vspace{-2mm}
\centerline{\includegraphics[width=13cm]{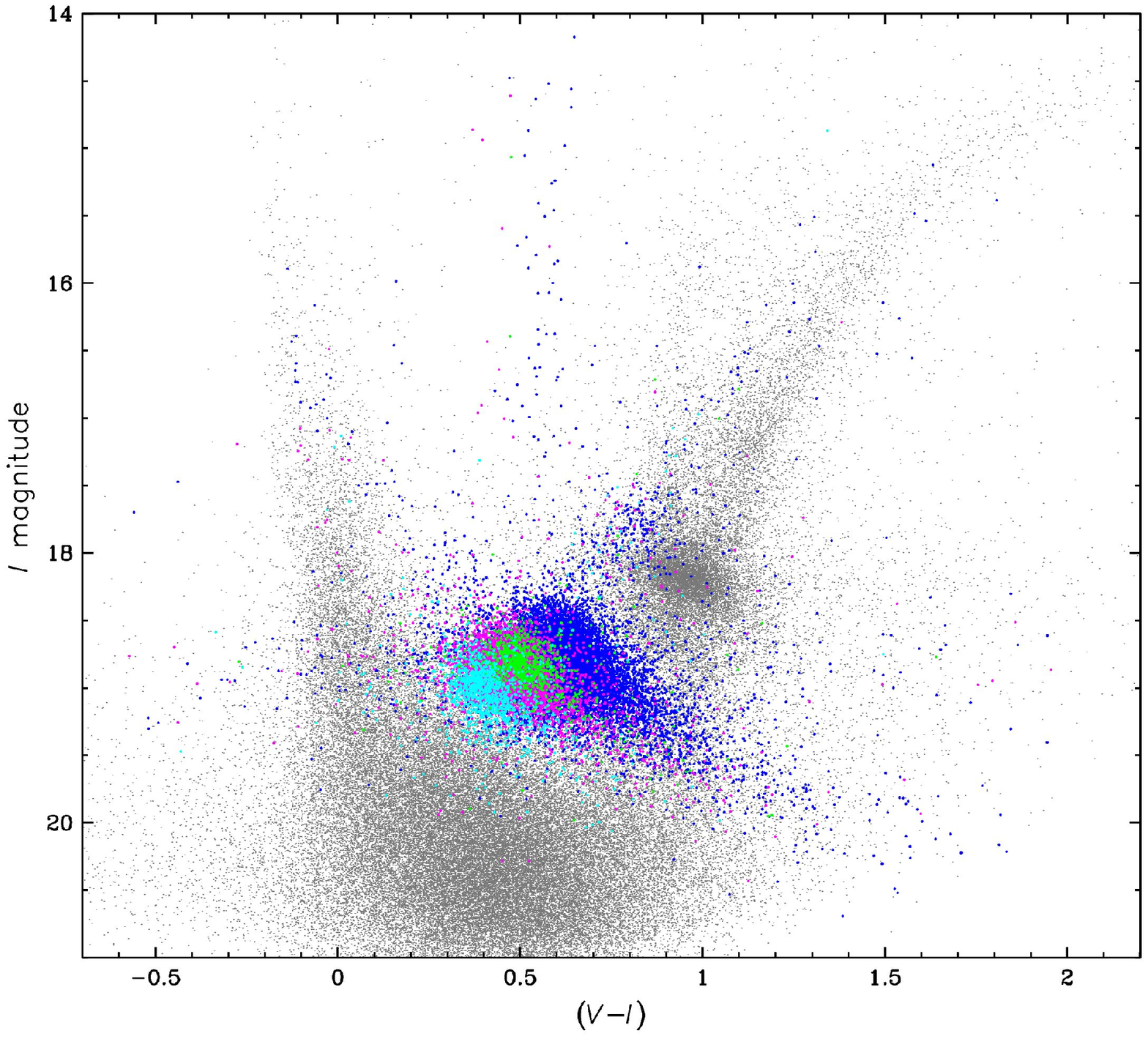}}
\FigCap{Color--magnitude diagram for RR~Lyr stars in the LMC. The color
symbols are the same as in~Fig.~2. In the background stars from the
subfield LMC100.1 are plotted.}
\end{figure}

The apparent {\it V}-band magnitudes of RRab stars have the modal value at
\linebreak19.36~mag. The overtone RR~Lyr stars (RRc, RRd and RRe stars) 
are somewhat brighter in this filter -- the most preferred {\it V}-band
magnitude for these stars is about 19.32~mag. In the {\it I}-band
fundamental-mode RR~Lyr variables are on average brighter (18.78~mag) than
overtone pulsators (18.88~mag).

Among variables classified as RRab stars the shortest periods are below
0.3~days, the longest periods are around 1~day, but more than 95\% of
fundamental-mode RR~Lyr stars have periods in the range of
0.45--0.75~days. Mean period of RRab stars in the LMC is $\langle
P_{ab}\rangle=0.576$~days, and it is close to the most preferred period
(Fig.~1) for these objects: 0.580~days. The periods of RRc stars range from
0.25 to 0.49~days with a mean value of $\langle P_{c}\rangle=0.337$~days,
and a modal period of 0.341~days. The first-overtone periods of RRd stars
are between 0.33 and 0.44~days with a mean value of $\langle
P_{d}^{1O}\rangle=0.363$~days and the most likely period of
0.357~days. Finally, for RRe stars $\langle P_{e}\rangle=0.270$~days and
the most frequent period is 0.272~days.

\Section{Conclusions}
In this part of the OIII-CVS we presented a selection of almost 25\,000
RR~Lyr stars in the LMC -- almost three times larger than the largest
sample of RR Lyr stars published to date. This is the largest set of RR~Lyr
stars identified so far in any environment.

This huge sample provides a unique opportunity to investigate in detail
various statistical features of these objects, like relations between
observational and physical parameters of RR~Lyr stars (\eg Jurcsik and
Kov{\'a}cs 1996, Morgan \etal 2007), incidence rates of the Blazhko
variables, or spatial distribution of RR Lyr stars in the LMC. Long-term
OGLE photometry can be also used for examining secular changes in RR~Lyr
stars.

\Acknow{
We are grateful to Prof. W.A.~Dziembowski for critical reading of the paper
and clearing up the problem of pulsating stars in eclipsing binary
systems. We thank Drs. Z.~Ko³aczkowski, T.~Mizerski, G.~Pojmañski,
A.~Schwarzenberg-Czerny and J.~Skowron for providing the software which
enabled us to prepare this study.

This work has been supported by the Foundation for Polish Science through
the Homing (Powroty) Program and by MNiSW grants: NN203293533 to IS and
N20303032/4275 to AU.

The massive period search was performed at the Interdisciplinary Centre for
Mathematical and Computational Modeling of Warsaw University (ICM
UW),\linebreak project no.~G32-3. We are grateful to Dr.~M.~Cytowski for
helping us in this analysis.}

\end{document}